# Space Sustainability: The Global Commons

Angelina Frolova[1] · Sascha Nick · Jean-Paul Kneib · Emmanuelle David · Carolyn Crichton · Sebastian Belanger Villanueva ·Mathieu Udriot · Marnix Verkammen

*École Polytechnique Fédérale de Lausanne (EPFL), Switzerland*

[1] Corresponding author: angelina.frolova@epfl.ch

## Abstract

Space is no longer a distant, barely reachable frontier of scientific exploration; it has become a critical infrastructure for life on Earth and a rapidly expanding zone of intense commercial activity. Yet this expansion is unfolding in ways that replicate, and amplify, the unsustainable dynamics already driving ecological and social crises on our planet. This chapter explains the concept of space sustainability as a multidimensional challenge; it raises the questions of governance, knowledge, power, justice, and global responsibility. It highlights both the profound benefits derived from space activities and the uneven distribution of costs and risks, from orbital debris and atmospheric impacts to privatization of essential services. We therefore argue that it frames space as a global commons that demands stewardship rather than pure exploitation, cooperation rather than competition, and governance that prioritizes long-term wellbeing rather than laissez-faire approach. Sustainability in space is inseparable from sustainability on Earth.

This chapter invites readers, whether they are space administration and industry actors, policymakers, scientists, or simply engaged citizens, to see space not as a limitless frontier, but as a shared domain whose future depends on collective choices made today.

# 1.Introduction

**This chapter discusses the challenge and the necessity to maintain our enduring access to space as a key commons of humanity.**

The value humanity derives from space depends fundamentally on the accessibility of space as a shared resource. The availability of this resource, along with the continued functioning of space assets, has direct implications for life on Earth as countless terrestrial systems and industries rely on the smooth operation of space infrastructure: telecommunication networks, weather forecasting, natural disaster monitoring, navigation, and much more.

Beyond these practical concerns, space exploration itself deserves our attention. The scientific pursuit of understanding more about the cosmos is not merely a luxury; it is a fundamental expression of human curiosity and purpose. As Carl Sagan (1980) beautifully observed, *"We are a way for the cosmos to know itself."* (Sagan et al., 1980, 00:06:26). To close the door to space would not only jeopardize the systems that sustain our daily lives but also betray one of the deepest meanings of human existence: the drive to explore beyond the known.

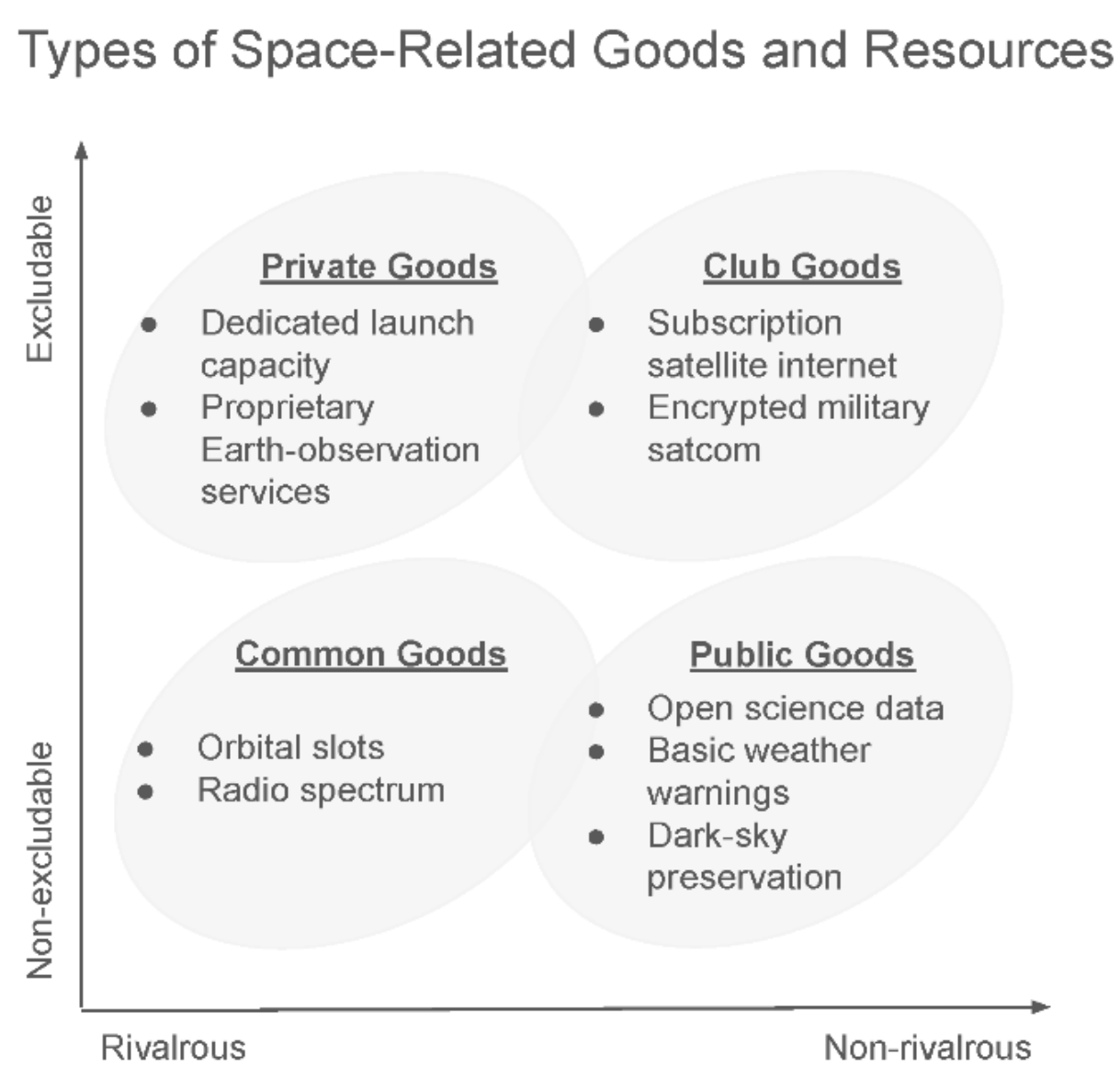

**Figure 1.1:** Types of space-related goods and resources, classified by excludability and rivalry in consumption. In reality, most things sit on a continuum; for example, satellite internet becomes rivalrous when capacity is reached, and orbital slots become excludable with suitable governance.

To act sustainably in space, we must see it as a commons (Figure 1.1). where some benefits of space can be privately controlled or restricted and others are broadly shared (excludability), and access to the commons by some actors may reduce what is available to others (rivalry in consumption). These dimensions distinguish private goods, club goods, common resources, and public goods within the broader space commons.

Many complex systems that endure must solve a central paradox: this commons depends on shared resources, yet they evolve through decentralized action. Every launch adds value to the collective infrastructure of observation, experimentation, communication or exploration; yet each also imposes risks (collisions, congestion, signal interference, vision occlusion) that no one intends but all must bear. When individual actors pursue benefits while drawing on shared resources, the system can drift toward overuse, degradation - a dynamic often described as the tragedy of the commons. The sustainability of space therefore hinges on our ability to design governance of space commons as carefully as we design technologies.

## 2. Three-dimensional problem statement

The conceptualization of space sustainability is largely grounded in the working definition provided by the Committee on the Peaceful Uses of Outer Space (COPUOS) of the United Nations Office for Outer Space Affairs (United Nations Office for Outer Space Affairs [UNOOSA], 2021). Their definition[2] articulates four main principles: 1) temporality, understood as the ability to conduct space activities indefinitely, 2) equity, referring to the fair access to space for all actors, 3) the peaceful use of outer space, 4) the environmental preservation of outer space.

Translating these overarching principles into a workable analytical perspective, this chapter adopts a three-dimensional framework that captures how sustainability unfolds across different facets of space activity. This vision is inspired by the work of Wilson and Vasile (2023) and is structured around three pillars: **Sustainability from Space**, **Sustainability in Space**, and **Sustainability for Space**. Sustainability from Space perceives space 'as a service' to support on-Earth sustainability. Sustainability in Space regards space 'as a resource' whose long-term viability depends on the responsible stewardship of its environments. Finally, Sustainability for Space examines the impacts of space activities and assets on the Earth systems.

### 2.1 "Sustainability from Space" (Sustainability enabled by space)

*'Sustainability from space'* captures the idea that space-based capabilities provide essential services helping societies monitor, manage, and improve sustainability on Earth.

#### 2.1.1 Earth observation for climate, agriculture, and disaster management.

For much of human history, natural disasters were seen as unpredictable 'Acts of God' or 'Acts of the Elements'. Today, however, humanity has a synoptic 'eye in the sky'. Space-based infrastructure now contributes at every stage of a disaster cycle - ***before***, ***during***, and ***after***.

**Before disasters**, meteorological satellites provide early warning by detecting cyclones, atmospheric rivers, heatwave development, wildfire risk, volcanic ash clouds, and rainfall systems. **During disasters**, satellites deliver real-time situational awareness. Optical and thermal imagers map wildfire fronts, lava flows, storm structure, and flood extent. Synthetic Aperture Radar (SAR) pierces cloud cover. Navigation constellations (GPS, Galileo) support evacuation routing, emergency communication timing, rescue coordination, and positioning of response teams. **After disasters**, radar and optical satellites map damage, identify collapsed buildings, track inundation and landslides. Combined with navigation data, these datasets guide aid logistics, assess infrastructure losses, inform recovery planning (UNOOSA, 2025a). In addition to aiding in natural disaster management, satellite remote sensing can provide other non-obvious benefits, such as, for instance, informing malaria risk models (Kazansky et al., 2016).

[2] *"The long-term sustainability of outer space activities is defined as the ability to maintain the conduct of space activities indefinitely into the future in a manner that realizes the objectives of equitable access to the benefits of the exploration and use of outer space for peaceful purposes, in order to meet the needs of the present generations while preserving the outer space environment for future generations."* (UNOOSA, 2021,p.2)

Yet satellites do far more than help societies during their worst moments. The very same Earth-observation systems that monitor natural disasters and disease spread, also support essential everyday needs. A notable example is **agriculture**. In the early days of Earth observation from space, using LANDSAT imagery, analysts were able to measure crop areas with a degree of consistency that had previously been impossible (Bauer et al., 1979). Others successfully used satellite-derived reflectance data to estimate crop water requirements (Estes et al., 1978) and track vegetation conditions across the African Sahel, producing early warning of drought and contributing to famine prevention (Tucker et al., 1983). Over the years, these capabilities matured. Already in the 2010s, high-resolution constellations such as Planet's nanosatellites were offering near-daily field-scale imagery, providing detailed maps of vegetation health at the level of individual parcels (Houborg & McCabe, 2016).

The two issues discussed in this chapter - extreme weather events and threats to food security - are expected to grow more severe as **climate change** alters weather patterns and disrupts ecosystems (IPCC, 2022, SPM §B.4.3; IPCC, 2023, SPM §B.1.3). That is why a discussion on *sustainability from space* would have been incomplete without a closer look at how satellites help us monitor the climate itself.

To appreciate how indispensable this vantage point is, it helps to do a simple thought experiment. Every few years, the world's scientific community assembles the most comprehensive assessment of climate knowledge, a massive international effort to synthesize thousands of studies into a unified picture, known as the Intergovernmental Panel on Climate Change report (IPCC, 2023; IPCC, 2022). If we removed every conclusion from a recent IPCC report that uses satellite observations, *what would remain of modern climate science?*

We would lose important global-scale longitudinal evidence for change in land cover and vegetation, ocean productivity, evapotranspiration and soil moisture, snow-cover duration, flooding, groundwater storage, forest condition, and coral-reef heat stress (IPCC, 2022, Ch. 2, pp. 209, 273; Ch. 3, p. 450; Ch. 4, pp. 565, 568, 570, 573, 581–582; Ch. 10, pp. 1514–1516; Ch. 15, p. 2056). Projections of future warming and sea-level rise would be weakened and multiple IPCC's high-confidence conclusions would become less certain (IPCC, 2022, Ch. 2, pp. 210, 293). Entire datasets central to monitoring climate extremes would be missing (e.g., MODIS). In short, without satellite observations, our climate knowledge would be a shell of its current self.

Ultimately, whether managing disasters, securing food supplies, or tracking climate change, space-based systems give us a vantage point from which the hidden becomes visible and the unpredictable becomes, at least to a certain extent, manageable.

### 2.1.2 Navigation and communication services.

Navigation and communication satellites have been central to the development of space technologies. In 2025, of the 12,812 active payloads, more than 90% were dedicated to communication or navigation functions (Data extracted from ESA DISCOSweb tool). OECD (2022) estimates that the present value of long-term global economic activity dependent on satellite navigation or telecommunication reaches around $190 trillion.

The concept of using satellites to determine precise geographic locations originated with the observation of Sputnik-1 from the Naval Center for Space Technology, using radio-frequency emissions (Alexandrow, 2008). Shortly thereafter, programs led by the U.S. initiated the development of what ultimately became the Global Positioning System (GPS). A functioning Global Navigation Satellite System (GNSS) requires ground control stations, Earth-orbiting satellites equipped with atomic clocks to ensure precise timing signals, and receivers capable of processing those signals into accurate position estimates (Alexandrow, 2008). Over the decades, technological advances have enabled miniaturization, sophistication, and digitalization of those components, reducing costs and enabling widespread commercialization. By 2024, Positioning, Navigation and Timing (PNT)-enabled services and device revenues accounted for $308 billion of the space economy (Pleney, 2025). In addition to the U.S. GPS constellation, sovereign GNSS networks were developed by Europe (Galileo), China (Beidou), and Russia (GLONASS).

Early systems were placed in Geostationary Orbit (GEO) at about 35,786 km because a single satellite could provide continuous coverage to a broad region. Over time, however, the long signal travel distance created latency and capacity limitations, and congestion in GEO became a growing vulnerability (Al-Hraishawi et al., 2023). These constraints prompted the development of Non-Geostationary Orbit (NGSO) systems: Medium-Earth Orbit (MEO), Low-Earth Orbit (LEO), and Highly Elliptical Orbit (HEO). They offer lower latency and more flexible coverage despite requiring larger constellations (Butash et al., 2021). As a result, a proliferation of communication satellites, particularly in LEO, has transformed the current space environment, including Starlink and Amazon Kuiper mega-constellations.

Both communication and navigation satellites are now essential to multiple Earth-based applications, and contribute directly to Sustainable Development Goals (SDGs) (Wood et al., 2024). Navigation systems support efforts such as wildlife monitoring, agricultural optimization (see Section 2.1.1), and epidemiological forecasting, while communication satellites facilitate information exchange, remote education, disaster response, and global connectivity (see Section 2.1.3). The increasing social reliance on these systems highlights critical vulnerabilities and calls for better network resilience. As mega-constellation deployments scale, the risks of orbital congestion and conjunction events increase (see Section 2.2.1), threatening industries and public services dependent on uninterrupted satellite-based communication and navigation.

#### 2.1.3 Global connectivity bridging digital divides.

Nearly three-quarters of the global population is now online, yet this leaves 2 billion people offline with stark disparities across income levels and regions (Sector, 2025; ITU, 2025). High-income countries achieve 94% internet penetration, while low-income countries reach only 24% (ITU, 2025). Moreover, connectivity does not guarantee quality access. A gender divide of 9% persists worldwide (ITU, 2022). Connectivity encompasses 3G-5G networks, which enable enhanced services, including telemedicine, remote education, and smart agriculture, thereby accelerating socioeconomic development in underserved regions. Space infrastructure plays a central role in advancing sustainability, citizen wellbeing, and access to essential services.

Bridging this gap requires multi-level initiatives (WEF, 2023). The primary challenge lies in infrastructure investment, requiring USD $1.6 trillion by 2030, and the substantial financing gap between nations (ITU, 2025). International organizations are mobilizing efforts to address this divide. The ITU has established programs including the Digital Infrastructure Investment Initiative with seven major development finance institutions, the Partner2Connect Digital Coalition for universal connectivity in least developed countries and small island states, and the Giga Initiative to connect every school to the internet by 2030 (ITU, 2025). Core infrastructure gaps and unclear digital policies remain significant barriers.

Space-based satellite connectivity offers solutions for remote areas lacking terrestrial infrastructure. In January 2025, OneWeb satellite technology was deployed in rural South Africa, initially supporting voter enrollment in the Northern Cape and Karoo regions (Satellite Today, 2025). This deployment illustrates both opportunities and complexities. While satellite connectivity enables civic participation and essential services, the same infrastructure serves commercial sectors including mining, oil and gas, and border security (Satellite Today, 2025). This dual nature demonstrates how connectivity infrastructure, while addressing gaps in underserved regions, simultaneously serves extractive industries and surveillance applications, which raises critical questions about equitable benefit distribution.

Failure to achieve universal connectivity risks economically excluding billions of people and jeopardizes SDGs. Education gaps widen as children in remote areas lack online learning access. Healthcare disparities deepen. Infrastructure vulnerability creates geopolitical and resilience risks (WEF, 2025). Rural communities cannot access financial services or employment opportunities, resulting in social exclusion and reduced civic participation. Additionally, deeper digitalization facilitates surveillance by governments, corporations, and threat actors, compromising freedoms and trust while potentially polarizing societies.

The Pact for the Future demonstrates commitment to bridging the digital gap across generations (UN, 2024). This requires multi-stakeholder cooperation among government, private sector, and civil society. Access to emerging services must benefit all, necessitating careful governance to ensure connectivity serves public good alongside commercial interests.

### 2.2 "Sustainability in Space" (Sustainability of the Space Environment)

*'Sustainability in space'* focuses on preserving the orbital environment and thereby ensuring that space remains accessible for future generations.

### 2.2.1 Space debris and orbital congestion

It is common to think of space as an endless expanse. Whether the Universe is truly infinite remains an open question suspended between mathematics and metaphysics, but one fact is already clear: the region of space closest to Earth is not limitless. Every object we send into orbit either remains there, re-enters the atmosphere, or has to be disposed of otherwise. It does not disappear into thin… *vacuum.* With current trends, we risk a situation in which near-Earth space becomes saturated with debris, making access to orbit (and, by extension, many of the systems we rely on) dangerous or even impossible. Ironically, *we may run out of space in space.*

This abstract concern becomes very concrete when looking at how Earth's orbit has actually filled over time. From lonely Sputnik-1 in 1957, the orbital population went to tens of thousands of cataloged objects. Collectively, these objects amount to more than 15,100 tons of material circling the Earth (ESA, 2025).

But the key question is not simply *how many objects are in orbit*; it is whether the environment can safely absorb the consequences of those objects being there. This leads to the idea of orbital thresholds - critical levels of crowding. In the 1970s, NASA scientists proposed that if a certain density of objects were reached in LEO, one collision would trigger another, creating a runaway cascade of debris (Kessler & Cour-Palais, 1978). This concept is now known as the ***Kessler syndrome***. Today, space agencies do not evaluate this scenario using a single tipping-point number, but rather use risk-based metrics that assess how close the environment is to collapsing. For example, ESA utilizes an Environmental Index (ESA, 2025, p. 138).

The collision hazard is the most evident risk posed by space debris: not only the possibility of catastrophic break-ups following major impacts, but also the cumulative damage caused by millimeter-scale fragments (NASA, 2012). Understanding the severity of these threats, spacecraft operators face an increasing collision-avoidance burden. Thousands of avoidance maneuvers performed annually cost operators money, fuel and operational lifetime. The lesser-known risks associated with a congested orbital environment are optical occlusion, which hampers our ability to observe the skies from the ground (see Section 2.2.2), and radiofrequency congestion which increases competition for limited spectrum.

*But should any of this come as a surprise?* Hardly. We have known that space debris will be a problem since the dawn of the Space Age. For example, in the 1960s astronomers were explicitly complaining that projects like West Ford[3] were 'cluttering space', and worrying about long-term interference with astronomy (Lovell & Blackwell, 1962; The Harvard Crimson, 1963).

*If we have long understood that orbital congestion is a threat, why have we not acted on it earlier?* This is a classic case of economic externality. Any operator investing in debris-limiting design or fuel reserves for disposal, bears the full expense, while the benefits (the slightly cleaner orbital environment) are shared by all, including competitors who contribute nothing. This creates a powerful disincentive for early movers. As a result, debris generation has been recognized in principle, but economically rational to ignore in practice for decades.

[3] West Ford was a 1961–63 U.S. experiment that released hundreds of millions of copper dipole needles into low Earth orbit to create an artificial radio-reflective belt for military communications.

The *Cosmos–Iridium collision* was a wake-up call. In 2009, the active satellite *Iridium-33* (≈560 kg) collided with the defunct satellite *Cosmos-2251* (≈900 kg), instantly obliterating both spacecraft (Pardini & Anselmo, 2017; Jakhu, 2010). Even though it happened at an altitude of 790 km, the debris clouds dispersed from 200 km to 1700 km, which means that one accidental collision affected almost the entire Low Earth Orbit (Liou & Shoots, 2009).

Now, major space agencies encourage end-of-life strategy consideration early in the design phase (NASA, 2013; ESA, 2023b). For satellites in Geostationary Orbit (GEO), the solution is the graveyard orbit, where spacecraft are maneuvered at the end of their functional life, safely out of the way. In LEO, however, where most contemporary satellites reside, this approach becomes unworkable. Small satellites dominating the LEO landscape are designed to minimize mass and cost; many have no propulsion at all. As a result, the only practical disposal method for LEO satellites is deorbiting, that is - sending them back into Earth's atmosphere. Large spacecraft are typically brought down via *controlled reentry* into designated oceanic regions, whereas the vast majority of smaller satellites undergo *uncontrolled reentry*, where they burn up in the upper atmosphere (see Section 2.3.3). Recently, Active Debris Removal (ADR) has emerged as a complement to disposal efforts. For example, ClearSpace aims at setting a precedent for cleaning up large, long-lived debris objects from orbit (ClearSpace, 2025).

Yet these initiatives alone cannot keep pace with the accelerating use of space (see Section 4 for a more detailed discussion). The cosmos may stretch beyond our comprehension, but the part of it we actively use is limited and already under pressure.

#### 2.2.2 Dark and quiet skies

Dark and quiet skies are defined as conditions where artificial light and spectral noise remain close to the natural background (IAU, 2020). Efforts to preserve sky quality date back over a century, with concerns about light pollution emerging alongside electric lighting in the late 19th century and spectrum management developing in the early 20th century to address conflicts among wireless services (UCDO, 2020; ITU-R, 2025). This protection falls within the global commons because light and noise from satellites cross borders and require coordinated governance. Governance, however, is uneven: responsibility for light pollution is distributed across environmental, spatial planning, and conservation agencies (Kyek, 2023), while radio spectrum is governed centrally through the International Telecommunication Union Radiocommunication Sector, which assigns frequencies and sets technical rules.

However, these long-standing governance structures are now under increasing strain. Recently, the cumulative effects of thousands of satellite constellations have increased diffuse sky brightness by reflecting sunlight from metallic surfaces and solar panels, reducing contrast and obscuring faint celestial features (Koller & Thompson, 2020). The International Astronomical Union Centre for the Protection of the Dark and Quiet Sky (IAU CPS) now leads advocacy and technical guidance, while groups such as the Committee on Radio Astronomy Frequencies (CRAF) analyze and monitor interference and help ensure protection for spectrum used by European observatories (CRAF, 2025). At the same time, the UN-level regulatory body COPUOS provides an institutional framework for discussing and potentially extending protection of dark and quiet skies under international space law (COPUOS, 2025).

Effects on optical astronomy experiments already illustrate the scale of the problem. The collective reflectivity and dense orbital population of satellites scatter sunlight in ways that significantly disrupt observations, particularly at altitudes between 500 and 1,200 kilometers where satellites remain sunlit for long periods during twilight (Bassa, Hainaut, & Galadí-Enríquez, 2022; Osborn et al., 2022). Following the first Starlink launches, between 1,600 and 2,000 satellites were visible above the horizon at any time, with 260 to 300 illuminated near twilight. These satellites are creating bright streaks in up to 40% of wide-field astronomical images during twilight (Hainaut & Williams, 2020; Bassa, Hainaut, & Galadí-Enríquez, 2022). Automated analyses can detect trails in a substantial fraction of images (Stoppa et al., 2024). Collectively, these findings show that the rapid expansion of satellites in low Earth orbit poses an escalating threat to ground-based optical astronomy, particularly for wide-field surveys essential to time-domain and cosmological research.

Radio astronomy faces similar challenges. Traditional protections such as radio quiet zones and data flagging are no longer sufficient because the number and diversity of satellites now produce persistent interference, including unintended electromagnetic leakage that overlaps with scientifically important bands (IAU CPS, 2024). This interference has already been confirmed in experiments such as LOFAR, which detected unintended Starlink radiation in the 110 to 188 MHz range, and MeerKAT, which observed out-of-band signals from GPS, Galileo, and GLONASS that reduce sensitivity (Di Vruno et al., 2023; Engelbrecht et al., 2024). Mitigation strategies such as trajectory-based subtraction remove some contamination but require detailed satellite information from operators, and hardware variability limits the effectiveness of software-based methods (Finlay et al., 2023; Bassa et al., 2024). With tens of thousands of spacecraft planned, these interference trends pose serious risks to transformational science projects such as the Square Kilometre Array (SKA Observatory, 2025; IAU CPS, 2024).

Satellite visibility further erases humanity's shared view of the night sky and weakens cultural, spiritual, and ecological connections that have existed for thousands of years, from ancient navigation to mythology, ritual practices, and seasonal planning (Hamacher, De Napoli, & Mott, 2020; Marín, 2011). More than 80% of the world's population already lives under light-polluted skies, with the Milky Way no longer visible to over one-third of humanity (Falchi et al., 2016). Preserving dark skies is essential to sustain Indigenous star knowledge, community identity, and the shared human heritage rooted in our relationship with the cosmos (IAU CPS, 2024;

McDonald, 2025). Increasing light pollution also disrupts ecological processes that depend on predictable light–dark cycles, confusing nocturnal wildlife, disturbing circadian rhythms, and altering navigation and habitat use (Yavuz & Bektas, 2024). Yet environmental protections rarely account for orbital light pollution, leaving significant regulatory gaps (Abbas, 2025).

Although it is difficult to fully quantify the economic cost of these impacts, the scale is clearly significant. In astronomy, survey performance does not translate directly into operational expenses, yet estimates suggest that constellations could cost a single experiment, LSST, an additional $34.8M over its lifetime due to extended observing time required to recover losses (David, Hellmich, & Crichton, 2025; Barentine et al., 2023). Global estimates suggest that total light pollution, including terrestrial, may reduce ecosystem service value by approximately $3.4 trillion per year, around 3% of global ecosystem service value and GDP (Anderson et al., 2024). Without coordinated regulation, improved spacecraft design standards, and greater transparency from operators, dark and quiet-sky conditions will continue to degrade faster than they can be protected.

#### 2.2.3 Planetary and lunar environment preservation

Working toward the preservation of an extraterrestrial body is not merely an ethical responsibility, but also a scientific necessity: biological, chemical and mechanical contamination can compromise future research, alter significantly the extraterrestrial environment and undermine the confidence of discoveries about the origin of life and planetary evolution. Preservation efforts start well before the space mission takes off, during the conceptual and design phase, and extend until the mission's end-of-life.

They involve applying biological control measures (COSPAR, 2026). This implies maximizing the spacecraft's cleanliness to reduce any risk of contamination of other environments. Alternatively, every measure should be taken to guarantee the safety of any biological experiments brought to extraterrestrial environments, such as the mini-biosphere within the successful *Chang'e 4* lunar lander in 2018 or the container of tardigrades in the Bearesheet lander that experienced a hard-landing in 2019. Such commercially and scientifically prompted biological cargo sparks debate (Shahar & Greenbaum, 2020). Additionally, missions can be planned to minimize contamination risk, as well as mechanical changes to a celestial body by choosing appropriate trajectories and accounting for sufficient Delta-V margins. As such, any unintended collisions and other forms of unplanned demise can be avoided or managed.

The issue of space debris around celestial bodies, albeit not yet urgent, must still be carefully considered, especially for lunar missions. Indeed, the rapid economic push toward lunar missions from governmental and commercial players (NSR, 2022), have prompted the development of a number of studies on the lunar orbital environment (Guardabasso 2021; Kirkpatrick, 2024; Turchetto and Guardabasso, 2024). Notable recommendations include the provision of propulsion capabilities to lunar spacecraft to accommodate for the irregular lunar gravitational field and for future debris avoidance maneuvers, the creation of a forum for transparent communication on orbital maneuvers and an increase in spacecraft tracking capabilities.

The end-of-life strategy is equally important to plan for. Where Earth-orbiting satellites may have options to demise in the atmosphere or move to graveyard orbits, it is often much more vague what the best course of action is for spacecraft in the rest of the Solar System. Favoring the avoidance of biological contamination on bodies of high scientific interest, one may take an approach like the *Cassini* mission, which demised safely in Saturn's atmosphere, rather than risking contamination of the planet's potentially habitable moons (Edgington and Spiker, 2016).

Dedicated end-of-life guidelines for lunar missions remain scarce (Kirkpatrick, 2024). Recent technical frameworks, including those proposed by EPFL (Turchetto and Guardabasso, 2024), outline disposal pathways tailored to mission altitude and orbit type, ranging from controlled lunar impact and Earth reentry to graveyard or heliocentric orbits. For Lagrange-point orbits, for instance, long-term residence near L2 is discouraged in favor of controlled impact or heliocentric transfer. These guidelines also highlight considerations such as protecting sensitive lunar regions and ensuring proper passivation. In general, such guidelines are still being developed and will evolve as lunar space activities grow.

Ultimately, planetary and lunar environment preservation is about foresight, in an area still relatively untouched by human activities. By adhering to established guidelines, planning end-of-life strategies carefully, maintaining cleanliness, and preparing adequate Delta-V margins, irreversible damage to celestial environments can be prevented. In doing so, mission planners uphold scientific integrity and honor humanity's responsibility to protect the Solar System.

## 2.3 "Sustainability for Space" (Terrestrial Sustainability of the Space Sector)

*'Sustainability for Space'* concerns terrestrial environmental impacts generated by space activities, including those arising from industrial supply chains, launches, and spacecraft reentry.

### 2.3.1 Industrial supply-chain and manufacturing footprint

The footprint of the space sector on Earth begins long before rocket launch (see Section 2.3.2) or satellite reentry (see Section 2.3.3): it starts with the industrial supply chain and manufacturing processes that make the space infrastructure possible.

Different methods assess these impacts. Simple qualitative assessment matrices, like the MET (Materials, Energy, and Toxicity) and monocriterion assessments, like carbon or water footprints, are often used for strategic decision-making at the level of an organization. For assessment at product or system level, however, Life Cycle Assessment (LCA) has become the European standard as it is multicriteria (computed impact scores for several indicators) and multisteps (covering the entire life cycle of the product or service under study). LCA is defined with four iteratives stages: 1) goal and scope definition, 2) life cycle inventory, 3) life cycle impact assessment, and 4) interpretation (ISO, 2006a; ISO, 2006b).

Design choices made in the earliest phases of a mission or spacecraft architecture lock in most of its eventual environmental footprint (e.g., ISO 14006; UNEP Life Cycle Initiative). Europe has been working toward practical ways to bring life-cycle thinking upstream into conceptual design. This is where simplified LCA[4] comes in: a streamlined iterative assessment approach that reduces data, scope, and analytical detail while still providing robust insights to support early design decisions (Tormena et al., 2024).

Major knowledge gaps, impeding LCA, remain. The upper-atmosphere effects of launch and reentry emissions are still poorly constrained (see Section 2.3.3 and Section 2.3.2). Space systems also raise boundary questions unknown in conventional LCA: while in-orbit propulsion emissions have no terrestrial midpoint impact, the orbital debris risk must be included to reflect impacts occurring beyond Earth's surface. For data collection to build the life cycle inventory (LCI), ESA has developed and maintains a space-specific database built on ecoinvent. Meanwhile, primes and their suppliers are responsible for collecting detailed foreground data - the most time-consuming part of any study.

After interpretation of the results, ecodesign options can be explored at two levels: early in system *architecture* decisions (e.g., for a launch vehicle: how many stages, using which propellant type, providing which mass and volume capacity), or later to optimize the system and its environmental performance. For both, technical alternatives with similar performances must be available to be able to choose a solution with reduced impacts. LCA's multicriteria approach is important here to avoid *burden shifting*: when performing ecodesign optimization on subsystems, it is always recommended to verify that a local gain (e.g., a subsystem with reduced impacts), does not shift the issue elsewhere at system level (e.g., if this greener subsystem is heavier and thus requires more propellant for the launch). Additionally, a gain in terms of environmental impacts, if accompanied by a gain in cost, could lead to a *rebound effect*, detrimental to the overall environmental impacts of the sector.

This chapter focuses on assessing the impacts the space systems have on the environment and society (inside-out point of view, *single materiality*). But as with any other activities, the space sector can be impacted by the changing environment and social considerations (outside-in, *double materiality*). For instance, some launch pads are at risk from rising sea levels, and some raw materials are more difficult to source. The double materiality assessment helps organizations align long-term sustainability strategy and stakeholders' expectations (ESA, 2023b). It identifies both the issues that matter *inside-out* and *outside-in*.

[4] Simplified tools in the space industries are being developed, including OASIS (Orbital projects: Assessment of Simple environmental Impacts Software) by CNES and Scalian, the NEST (Navigating the Eco-design of Space systems and missions Tool) by François Cluzel, or the ACT (Assessment and Comparison Tool) by EPFL Space Center and its consortium, now maintained by EcoDeltaV Sàrl.

Together, these elements reveal two intertwined questions - *How do we impact the environment?* and *How the changing environment is impacting us?*

### 2.3.2 Atmospheric impacts from launches

One differentiating aspect of the space sector is that it is the only sector that systematically emits in all layers of the atmosphere. Regardless of whether a spacecraft re-enters the atmosphere (see Section 2.3.3), or exits Earth's gravitational well, it has to be brought up by a launcher that emits gas and particles. The scientific community still has a limited grasp on the exact effects of such emissions through the atmosphere (Miraux, 2022; Fischer & Fasoulas, 2024; Udriot et al., 2025). It is estimated that two thirds of a typical rocket propellant is burned above the tropopause (Ross, 2014), where emission effects are least understood.

Part of the knowledge gaps lies in the lack of precise understanding of the elements that are emitted. Indeed, while the propellant quantity and type are known (i.e., one knows what fuel and oxidizer is being loaded into the rocket), much research is still ongoing on the way it is burned and ejected as the launcher crosses all layers of the atmosphere. The so-called primary emissions, originating from the combustion and directly released at the exit of the nozzle are sufficiently understood through an ideal reaction, but do not paint the full picture. Secondary emissions (i.e., compounds that are created in the hot exhaust plume and in the surrounding atmosphere due to chemical reactions) are highly dependent on atmospheric conditions across altitudes. Such emissions include among others nitrogen oxides (NOx), nitrogen ($N_2$), water vapor formed by exhausted hydrogen molecules (H and $H_2$) that react with the atmosphere's oxygen molecules ($O_2$) and carbon monoxide (CO) that reacts in the exhaust plume to form carbon dioxide ($CO_2$) (Dominguez Calabuig et al., 2022). Moreover, launchers produce incomplete combustions that are difficult to estimate. Black carbon (often compared to soot) is one of the most important emissions and is estimated to have major effects on various impact categories. (Dominguez Calabuig et al., 2022).

Today, launch activity has a limited effect on ozone depletion, estimated at below 0.1% (Voigt et al, 2013), but this could significantly increase with space launch activity. Some studies suggest a significant impact by 2050 (Ross & Vedda, 2018; Miraux, 2022), however this estimate remains highly uncertain.

A special note should be made on reusable launchers, in particular those with propulsive reentry. While the production impacts are spread over multiple launches, the launcher has to carry more propellant for its reentry maneuver, causing greater propellant burn rates and thus emissions across its ascent profile. Additionally, the reentry burn results in a spike of emissions in the highest layers of the atmosphere, compared to non-reusable launchers' emission profile. Finally, the last braking and landing burn cause further low-altitude emissions. Besides this, one could also include the impact of the recovery and refurbishment. Overall, it is difficult to give a black-or-white comment on reusability in launchers as research on their impact is ongoing, but one ought to be critical about the common idea that launcher reusability equates to an increased launcher "sustainability" (Dominguez Calabuig et al., 2022; Miraux, 2022).

### 2.3.3 Reentry pollution

When a satellite or rocket re-enters the atmosphere, it effectively becomes a *human-made meteor*. The intense heat and friction cause the object to vaporize and shed material. This process creates a variety of atmospheric pollutants, which could have adverse environmental impacts.

A major concern is the *ozone layer.* Decrease in its depletion is one of the textbook examples of successful mitigation of human-induced ecological problems. Thanks to the actions taken under the Montreal Protocol, the ozone layer is recovering (WMO, 2022). However, nitrogen oxides from reentry heating (Ryan et al., 2022) and aluminum oxides catalyzing chlorine activation in the stratosphere (Ferreira et al., 2024) could threaten this progress.

Broader *climate* harm can come from the spent rocket bodies and defunct satellites: when burning up in the atmosphere, they leave a trace of metal vapors that condense into aerosol particles (Murphy et al., 2023). These particles accumulate in the stratosphere, where they absorb shortwave light and may thus contribute to the warming effect, similar to that of volcanic ash (Kapoor et al., 2025). Importantly, human-made objects now contribute *as much aluminum to the atmosphere as natural meteoroid ablation* (Ferreira et al., 2023).

And while aluminum is a major environmental concern, both due to its abundance in spacecraft materials and its newly recognized radiative activeness; it is not the only reentry-induced climate threat. Recent high-altitude sampling campaigns by NASA reveal that spacecraft demise injects a broad suite of metals in the stratosphere: titanium, chromium, nickel, iron, magnesium, copper, lithium (Murphy et al., 2023). Critically, we still remain in the dark about how these metals may affect our climate (Murphy et al., 2023; Schulz et al., 2025; ESA, 2023a). Transition metals such as chromium, nickel, and copper are known catalysts for heterogeneous chemical reactions, yet their behavior in the cold, dry stratosphere remains almost entirely unstudied. Titanium and magnesium oxides, which form at high temperature during reentry ablation, could influence aerosol composition and cloud nucleation in ways that current climate models do not capture. Moreover, researchers warn about the *scaling risk:* what is 'tolerable' now, can soon turn into a significant atmospheric burden once the projected increase in satellite numbers fully materializes (Maloney et al., 2025; Ross & Jones, 2022).

We are thus presented with a dilemma - a *space sustainability paradox* (Wilson & Vasile, 2023): space assets are crucial for multiple dimensions of human life on Earth, including sustainability (see Section 2.1.1), and maintenance of space assets requires riddance of defunct spacecraft (see Section 2.2.1), which can be done by cremating them in our atmosphere. Hence the paradox: ***We launch satellites to protect Earth. We burn them up in the atmosphere to protect space. In the process, are we damaging the planet we intended to help?***

An economist might argue that the answer to this paradox lies in a cost–benefit calculus: if the societal, environmental and economic value generated by space-based services is tangibly higher than the marginal pollution created by satellite burn-up during reentry, then the trade-off may be considered acceptable. Yet this calculus is not straightforward. While benefits of space exploitation are evident and abundant, the research on the environmental cost is still scarce (Murphy et al., 2023; Schulz et al., 2025; ESA, 2023a).

*What, then, should we do with satellites once their missions end?* The dominant policy answer today is clear: reentry is the default disposal pathway for LEO spacecraft (UNOOSA, 2010; ESA, 2023a; NASA, 2019; ISO, 2023; ECSS, 2024). NASA requires that the probability of human casualty from any re-entering object be below 1 in 10,000 (NASA, 2012). One way to meet this threshold is to ensure that the spacecraft does not survive reentry in any meaningful form, an approach called **Design-for-Demise**, where materials and components are intentionally selected to fragment, burn, or vaporize before reaching the ground. From this perspective, "demisable" spacecraft are not only safer but seemingly more sustainable: nothing hits Earth and nothing remains in orbit. Yet this is only one side of the discussion. As Ott and Bonnal (2025) argue, the same design choices that make a satellite safer upon reentry, may make it leave more ablated materials in the atmosphere. In parallel, designing for total demise eliminates the possibility of post-mission reuse or material recovery, reinforcing a linear *'launch–use–discard'* model rather than a circular one. This counterview is called **Design for Non-Demise**.

And so the dilemma remains unresolved. Should we design spacecraft to vanish or to come back? For now, the two discourses coexist. The choice here is less a technical question than a question of what kind of space economy and environmental ethics we are prepared to build.

### 2.4 Summary of the Three Dimensions of Space Sustainability

The three pillars of space sustainability introduced by Wilson and Vasile (2023) reflect different relationships between the space sector, the orbital environment, and Earth's ecosystems. Each pillar is associated with distinct market dynamics that can either reinforce or undermine sustainable development. Wilson and Vasile (2023) also point out that two pillars - 'Sustainability in Space' and 'Sustainability for Space' can be merged into one: 'Sustainability of Space' - a unified lens that captures the orbital externalities and terrestrial environmental impacts. Table 2.4 summarizes these pillars and their implications for business decision-making within the broader tension between sector growth and the stewardship of space as a shared commons.

**Table 2.4** Three dimensions of space sustainability

| | Sustainability from Space | Sustainability of Space | |
|---|---|---|---|
| | | **Sustainability in Space** | **Sustainability for Space** |
| **Definitions** | Sustainability enabled by space | Sustainability of the space environment | Terrestrial sustainability of the space sector |
| ***Examples*** | *Earth observation for climate monitoring, disaster response, navigation and timing systems* | *Space debris mitigation, collision avoidance, traffic management, radio spectrum congestion and interference prevention* | *Industrial supply chain footprints, rocket launch emissions, atmospheric pollution from spacecraft reentry* |
| **Market dynamics** | Strong public-good value but often limited direct profitability. Markets tend to underprovide these services without public programs. | Individual operators benefit from orbital access but bear only a fraction of the collective risk. | Environmental costs are largely unknown and externalized to ecosystems, while benefits accrue to service providers. |
| **Implications for business decision-making** | Target market choice: high-profit (telecom, broadband, proprietary EO analytics) or public-good (climate monitoring); data governance model: proprietary/vertically integrated or open/shared ecosystems. | Number of satellites, orbital regions occupied, investments in debris mitigation (e.g., propulsion systems and end-of-life disposal capabilities), operational risk tolerance (e.g., collision avoidance strategies). | Launcher type and cadence, manufacturing strategy (e.g., mass and materials), satellite lifetime (longer-lived satellites or rapid replacement cycles), end-of-life design (e.g., design-for-demise, controlled reentry). |
| **Implications for governance** | Public institutions often need to subsidize or anchor markets for high-value sustainability services (e.g., climate monitoring programs). | Classic global commons dynamics: without coordination mechanisms, firms face incentives to expand faster than the orbital environment can safely sustain. | Policy frameworks need to introduce launch regulations and life cycle assessment requirements. |
| **Growth-sustainability tension** | Satellite broadband, commercial EO and geospatial analytics markets see rapid expansion. These markets scale with more satellites, higher revisit rates and proprietary data streams, while sustainability often requires open data and public-good missions. | Each firm benefits from launching more satellites, occupying orbital regions early and increasing redundancy; but collectively this leads to orbital congestion, increased collision avoidance costs and increased cascading debris risks. | Scaling of the space industry implies increased launch cadence and mass production of satellites. Many environmental impacts occur outside of the market transaction: they remain externalized and unpriced; firms have limited economic incentives to minimize them. |

In practice, the three categories overlap and are interdependent. Activities that enable sustainability on Earth rely on space infrastructure, whose deployment and operation affect both the orbital environment and terrestrial ecosystems. For example, expanding satellite constellations helps climate monitoring and connectivity, yet it aggravates the debris risk and atmospheric impacts from launches and reentry. Actions in one pillar inevitably generate consequences in the others, linking benefits and externalities across system boundaries.

## 3. Risks and benefits of space activities

Space activities generate remarkable scientific, cultural, economic, and security benefits. Almost all associated costs, from debris to mining and atmospheric pollution, are collective, while several of the most problematic activities privatize gains. Strikingly, the space uses with the greatest public value, such as Earth observation and fundamental science, tend to have comparatively small adverse footprints, whereas high-impact commercial deployments create disproportionate risks. This section classifies seven categories of space activities to illustrate the tension between public-good benefits and negative externalities, and builds the foundation for governance principles, drawing on Elinor Ostrom's work (see Section 4.4).

### 3.1 Understanding the Universe

The observation of the Universe over the last centuries has been essential to the discoveries of the fundamental laws of physics. Today, observations of the Universe continue with space mission exploration of the Solar System and with space telescopes observing at wavelengths blocked or impacted by the Earth atmosphere. Since their establishment, NASA and ESA have each launched dozens of science missions, totaling hundreds of spacecraft (Parmar et al., 2025.; NASA, 2017). Most of these missions are not in LEO and do not compound the Kessler Syndrome (see Section 2.2.1), but they of course generate atmospheric pollution.

Their climate/biodiversity benefits, however, are very strong, including major contributions to the understanding of planetary systems, climate models, and Earth system science. The inclusion and the wellbeing impact is generally positive, but depends on equitable access to research results and participation in scientific missions. Most astronomy missions provide open access to final data and results in allowing any scientist to benefit from the outcome of the astronomy mission. For planetary exploration, open access of data is sometimes more constrained, but results are generally open. A possible risk in the future is the privatization of scientific missions and findings in the case of shortage of public funds.

### 3.2 Shared destiny and human meaning

With the early exploratory space missions, a more global understanding of our place in our Universe has emerged. The iconic "Earthrise" on the moon's horizon taken in 1968 and the "Blue Marble" taken in 1972 have profoundly shifted human perspective, emphasizing Earth's unity and vulnerability. They have galvanized the modern environmental movement, leading to initiatives like the "Earth Day"(Poole, 2008). Voyager-1, launched in 1977, passed most of the planets of our Solar System and is now close to reaching the 1 light-day distance, demonstrating the challenge of interplanetary travel (the closest star to the Sun is 4.3 light-years away) (NASA, 2024). Similarly, space telescope images (Hubble and James Webb) have revealed spectacular pictures of our Universe inspiring education and strengthening planetary stewardship.

### 3.3 Monitoring earth's climate, biodiversity, and societies

Earth observation (EO) provides high-resolution monitoring of vital Earth systems, including forests, water cycles, land use, sea-level, greenhouse gas (GHG) emissions, glaciers, urbanization, and weather forecasting. This data is essential for critical global efforts, supporting climate policy frameworks like the Paris Agreement, biodiversity tracking required by the Convention on Biological Diversity (CBD), disaster response coordination, and agricultural resilience planning (see Section 2.1.1).

On the flipside, the expansion of EO capabilities introduces environmental risks, including orbital debris, rocket emissions, and atmospheric pollution. While the public satellites have an open sharing policy of their data, the commercialization of the private EO observations does not provide equitable distribution of the EO data.

The current impact on climate and biodiversity is overall strongly positive, but this benefit may be undermined if access to the data is restricted; consequently, the impact on inclusion and wellbeing is mixed. To mitigate these issues, we need shared governance, fair data access, establishment of global environmental data commons, maintaining/developing public investment in open-data satellite programs (e.g., European *Copernicus* program).

### 3.4 Navigation and communication systems

Navigation satellites are broadly used nowadays, providing substantial benefits for our society, for example, monitoring of infrastructure, shipping logistics, aviation safety, and emergency response systems (see Section 2.1.2).

The deployment of large constellations (e.g., Starlink) has intensified concerns about debris generation, frequent launches and re-entries. Unlike the geostationary regime, where the International Telecommunication Union (ITU) coordinates both orbital slots and radio spectrum to ensure fair access, LEO remains governed primarily through national licensing systems (ITU, n.d.; Sun & Liu, 2025). As a result, early entrants effectively shape and occupy valuable orbital shells, which creates risks of market dominance, reduced affordability, diminished sovereignty for later users. Because LEO orbital resources are finite, the absence of an international allocation framework poses long-term sustainability challenges.

### 3.5 Technology development: spillovers and risks

Technology innovations have been essential for space development, because of the inherent constraints of the space environment, such the need of light and robust materials, miniaturization of electronics, limited energy budget, limited communication bandwidth, harsh space environment (high level of radiation, high temperature variation, vacuum). All these challenges have pushed for the development of innovative solutions that have been positively used in other domains on Earth. Yet, many of the space technologies have been captured by commercial actors while the "space" costs (in terms of pollution, GHG emissions, space debris, material extraction) are borne by society. Other risks include misuse of advanced technologies (dual-use AI, surveillance, weapons guidance systems), and unequal access to innovation.

It is important that space innovation (funded by taxpayers) is conducted in a better transparency, using open licensing for a wide dissemination. Technologies with public benefit should be prioritized and businesses heavily relying on space activities (e.g., LEO telecom constellations) should be subject to a social license in order to operate.

### 3.6 War, defense, and militarization

Since the beginning of access to space, the military have been using satellites for reconnaissance imaging, early warning, monitoring of various assets, and telecommunications. In many cases, these space assets reduce uncertainty and can prevent conflict escalation.

Yet, militarization of space can be destabilizing and lead to serious conflicts or risks to the space infrastructure. For instance, the Anti-Satellite Tests (ASAT) where the military destroys its own satellite tests will lead to a cloud of smaller space debris which are harmful for numerous active satellites. Other examples such as proximity operations close to a military or civil spacecraft may lead to conflict escalations. Finally, cyber-vulnerabilities warrant increasing caution. It is important to not neglect offensive behavior, and develop global governance for cooperative early-warning systems and transparent conflict-resolution mechanisms.

### 3.7 Long-term futures: planetary protection and the 'Plan B' illusion

Planetary protection includes vital global safety research, as well as resource-intensive escapist efforts (such as establishing a Moon- or a Mars-based colony). It is a legitimate goal for humanity to protect its space assets from extreme solar flares, and eventually develop the capacity to deflect an asteroid.

Yet, the narrative of planetary colonization is largely techno-utopian, often diverting attention toward unfeasible problems, at least in any useful time frame. Still, it must be acknowledged that some associated technologies (e.g., closed-loop recycling systems for Mars colonies) may benefit society, provided that the focus is on *public benefits* and *open data/research policy.* The very notion that we have a *Planet B* risks complacency about Earth's urgent ecological challenges, since it implies that large-scale human settlement elsewhere is a viable contingency. Thus, political leaders must counter this narrative by emphasizing that space exploration should reinforce humanity's commitment to Earth, not serve as an alternative.

### 3.8 Conclusion and implications

The diverse benefits of space activities are undeniable, yet they remain inseparable from collective costs that are growing rapidly with the expansion of commercial and strategic uses of orbit. The cases above show that activities with the highest public value often have the lowest footprint, while profit-driven ventures introduce systemic risks that outpace current governance. Table 3.1 summarizes and Figure 3.8 illustrates who benefits and who bears the costs.

**Table 3.1:** Overview of key benefits and issues for each major space-related activity

| # | Benefits | Issues | Impact on: climate and biodiversity | Impact on: inclusion, wellbeing | Governance recommendations (inspired by Elinor Ostrom) |
|---|---|---|---|---|---|
| **1** | Understanding energy, matter, gravity, stars and planets, life | Minimal, mainly space debris | ++ | + | Establish international research frameworks ensuring equitable sharing of knowledge and non-exploitative use of findings |
| **2** | Culture, shared destiny and meaning "Earthrise"; inspiration to study science | | ++ | ++ | Promote global, inclusive participation in symbolic missions and shared narratives, fostering a sense of planetary citizenship |
| **3** | Monitoring the Earth: climate, weather, forests, agriculture, cities, GHG, glaciers … | Space debris, pollution, material extraction, unequal access, transfer of wealth from taxpayers to companies, privatization of critical resources and their governance | ++ | – / + | Use shared decision-making and fair data access to strengthen international agreements to act based on results |
| **4** | Navigation and communication | | – / + | – / + | Limit resource use and activity level; ensure equitable access to satellite networks; prevent monopolization |
| **5** | Technology development | Terrestrial issues dominate (e.g., control over technologies and misuse for profit or harm) | – / + | – / + | Limit resource use and activity level; enforce transparency; prioritize tech for public benefit (open licensing, knowledge sharing) |
| **6** | War and defense | Zero-sum game, terrestrial militarization extends to space, potentially end of humanity | – | – | Strengthen int'l space weapons bans; create joint conflict resolution mechanisms and shared early-warning systems |
| **7** | Very long-term: deflect major space collisions, live on other planets | Risk of ignoring urgent crises of biodiversity, climate, inequality under the illusion of "Plan B" | – | – | Ensure long-term research does not distract from today's crises |

The implications are clear. Space sustainability, like Earth sustainability, requires global coordination, transparency, equitable access, and regulatory frameworks that treat orbital space as a shared commons (see Section 4).

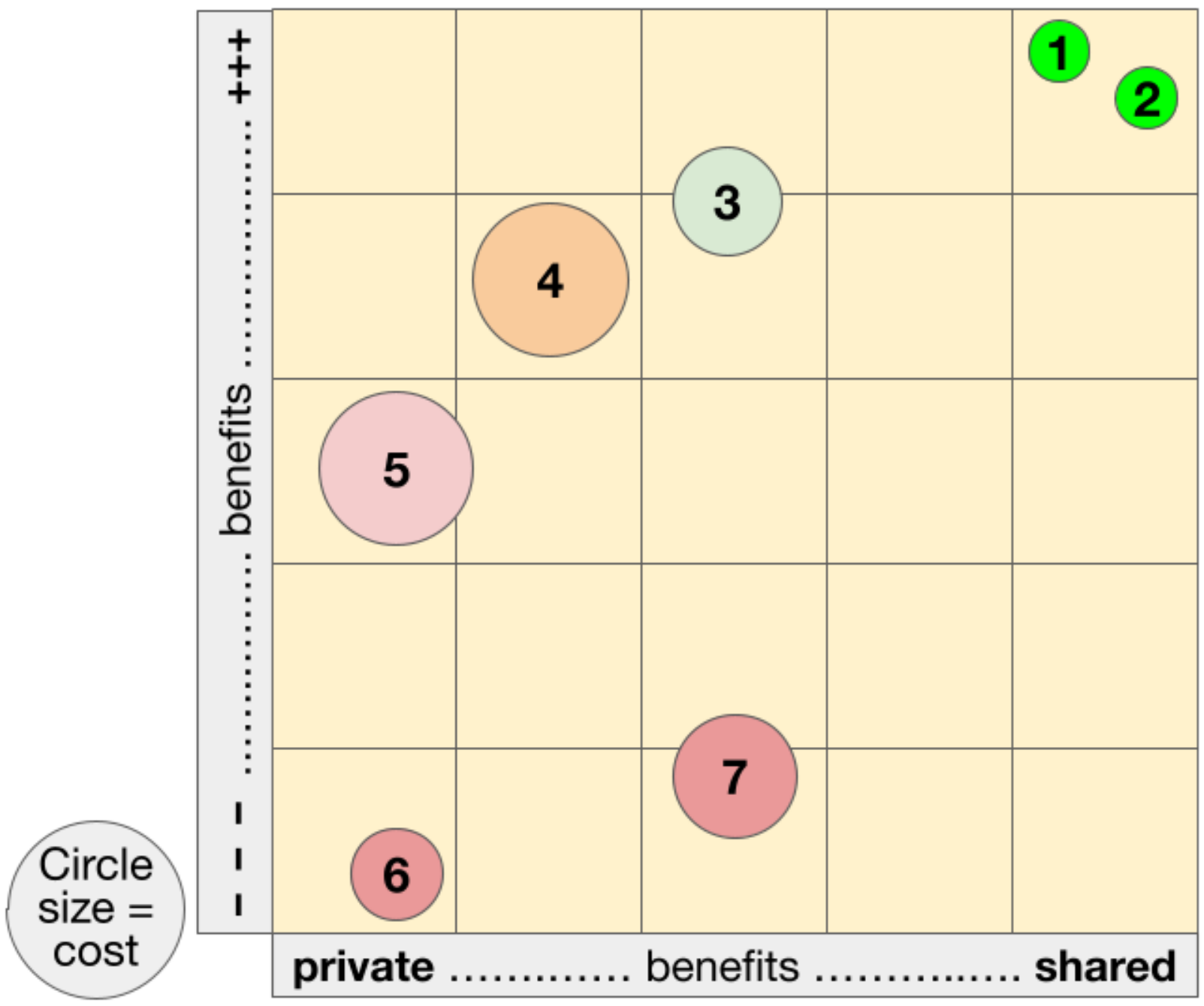

| # | Benefits |
|---|---|
| 1 | Understanding energy, matter, gravity, stars and planets, life |
| 2 | Culture, shared destiny and meaning "Earthrise"; inspiration to study science |
| 3 | Monitoring the Earth: climate, weather, forests, agriculture, cities, GHG, glaciers … |
| 4 | Navigation and communication |
| 5 | Technology development |
| 6 | War and defense |
| 7 | Very long-term: deflect major space collisions, live on other planets |

**Figure 3.8:** Distribution of benefits and environmental and social costs per space activity

# 4. Critical evaluation and future directions

## 4.1 Today's thinking, discourse and actions

Today's discourse often rests on assumptions such as techno-optimism (space technologies will bring broadly beneficial innovations), a focus on benefits (connectivity, climate data, economic growth) more than the environmental or social costs, a belief that private enterprise should be largely unconstrained because "innovation will trickle-down", and simultaneously a notion that actors must cooperate globally to manage the space commons.

> "*Given the unique nature of this topic, no single nation or entity can drive meaningful change without the coordinated action and cooperation of governments, satellite owner-operators or manufacturers, and astronomers from around the world*." (UNOOSA, 2025b, p.1)

> "*The growth in number of satellites also impacts amateur astronomers, and the general connection between humanity and the night sky, including for Indigenous communities*." (UNOOSA, 2025b, p.1)

> "*The agreement … ensures that Starlink … reduce[s] the optical brightness of their satellites ... via darkening, deflecting light away from the Earth, attitude maneuvering, or other ideas to be developed; maintaining orbital elevations at ~700 km or lower; and providing orbital information publicly …*" (IAU CPS, 2023)

The first quote shows a commitment to multilateral cooperation and mitigation, and quote two indicates growing concern about cultural, environmental and scientific costs, including the damage to astronomy, to universal access to the sky, and to intangible values such as the "night-sky heritage" for Indigenous communities. Yet quote three suggests that even when private companies attempt mitigation, the measures remain limited and subject to future "ideas to be developed". This is a weak commitment, without deeper reflection on the long-term sustainability of such constellations. In this sense, contemporary space-sustainability discourse often links bold visions to modest actions: many actors speak of cooperation, benefit and progress, but much less about the potential for irreversible environmental, cultural or social losses. Without translating discourse into effective regulation and inclusive governance, the gap between what is needed and what is protected will widen.

## 4.2 Conflicting logics and paradoxes

Throughout the sections, the reader will already have noticed several points where the pursuit of industry growth and the pursuit of sustainability reveal unresolved tensions. Whether discussing the environmental and economic dimensions of space activities, the risks and benefits of satellite operations, or the threefold structure of sustainability *from, in*, and *for* space (Wilson & Vasile, 2023), a recurring pattern becomes visible: space sustainability is not a single, unified objective, but a field shaped by competing logics which generate paradoxes when contradictory values or goals coexist and must be pursued simultaneously.

Table 4.2 Conflicting logics within the growing space domain.

| Logics | Conflicts |
|---|---|
| **1.1** Use more satellites to monitor climate, manage disasters, enable connectivity, support essential services on Earth. | **1.** The solution to Earth's problems increases pressure on the orbital environment. |
| **1.2** Limit orbital use to protect the fragile space environment from congestion and debris. | |
| **2.1** Deorbit satellites and defunct spacecraft to mitigate orbital congestion and debris accumulation | **2.** The solution to space problems increases pressure on the Earth's environment. |
| **2.2** Limit atmospheric pollution and casualty risk from spacecraft and debris reentry. | |
| **3.1** Build massive constellations with thousands of satellites to ensure continuous service even if assets fail. | **3.** Redundancy improves system performance but degrades system safety. |
| **3.2** Keep orbital population low to avoid collisions, interference, and Kessler-type cascades. | |
| **4.1** Maximize shareholder returns, market share, and commercial throughput. | **4.** Costs of sustainability are borne privately; benefits are shared globally. |
| **4.2** Preserve the orbital environment as a global commons for all current and future users. | |
| **5.1** Lower launch costs, rapid expansion of constellations, higher launch cadence, commercial scale-up, competitive advantage. | **5.** The same drivers that make space economically vibrant make the environment less resilient. |
| **5.2** Minimize traffic, manage debris, reduce launches, strengthen mitigation and disposal rules. | |
| **6.1** The Outer Space Treaty's principle of free access and use. | **6.** A legal framework built for an era of scarcity struggles with today's era of abundance. |
| **6.2** Growing recognition that actors must avoid harmful interference and cumulative environmental impacts. | |
| **7.1** Business models depend on near-term deployment, service delivery, and market capture. | **7.** Firms optimize for quarters; the orbital environment degrades over generations. |
| **7.2** Sustainability requires decades-long planning horizons and risk mitigation. | |
| **8.1** Explore, extract, and develop new missions and resources. | **8.** Ambitious exploration increases risks of irreversible biological or environmental harm. |
| **8.2** Avoid contamination of celestial bodies and biospheres. | |

Economic and financial dynamics introduce tensions that challenge the long-term sustainability of outer space activities. Democratization of space, mainly characterized by affordable launching capabilities and a higher technological maturity, has facilitated the proliferation of constellations, particularly in LEO. Space businesses can leverage private, public, and institutional capital to design large constellation networks, relying on redundancy rather than individual reliability. Latest developments have demonstrated strong interest for mega-constellation projects, driven by high profitability expectations. Yet, a more densely populated orbital environment naturally increases the probability of conjunctions events, threatening the financial resilience of the business models.

Table 4.2 shows that the space domain is defined by these multiple, overlapping paradoxes. Strengthening sustainability *from* space can weaken sustainability *in* space (see Section 2.2.1); stabilizing orbit can shift impacts back *to* Earth (see Sections 2.3.2 and 2.3.3); and commercial, technological, and geopolitical incentives often work at cross-purposes with long-term stewardship. These tensions coexist and reinforce one another, underscoring why space sustainability requires continuous balancing rather than definitive resolution.

### 4.3 Unresolved scientific and technological needs for a sustainable future

The previous sections have analyzed space sustainability in its three dimensions – *for, from* and *in* space. When we synthesize these three pillars back, a critical question emerges: what are we still missing in terms of knowledge and technology to enable space sustainability?

The first challenge is managing space assets to ensure the long-term use of outer space for present and future generations in an equitable way. This can be summarized as establishing governance frameworks, technologies, and models for space traffic management. We are missing clear *"rules of the road"* (Rossi et al., 2024) and the infrastructure to control and manage circulation (Bonnal et al., 2025) This is highly operational and must evolve quickly to adapt to emerging markets, not only in LEO but eventually extending to the Moon, Mars, and beyond.

Discussions unfold around active debris removal, in-orbit manufacturing, and recycling. However, business models are still being validated and markets have yet to mature. Furthermore, while standards are being developed, gaps remain in defining what needs to be done. The ESA's Zero Debris charter presents technologies and enablers needed to fulfill the vision of a sustainable space future (ESA, 2023). To select appropriate technologies and solutions, we must fill major gaps in understanding and modeling both the space and Earth environments. While we have known about the risk of the Kessler syndrome since the 1980s, simulations of its impact on the growing population vary considerably (Colombo et al, 2025). New research projects are investigating how space weather affects the debris population, particularly smaller objects. Large constellations have changed the paradigm: thousands of smart spacecraft now push the limits of what was considered maximum orbital capacity. The highly dynamic nature of orbits increases complexity, while accurate observation remains difficult. Challenges persist in limiting orbit propagation errors and ensuring model interoperability across different actors. When examining space activities' impact on Earth, LCAs are limited by a range of factors (see Section 2.3.1).

Key questions remain: how do we bridge the multifaceted nature of space activities, linking safety, security, and sustainability when these issues are currently addressed separately at national, regional, and international levels? How do we integrate sustainability impacts into costs, value assessments for investors, and insurance considerations? Interdisciplinary integration, such as through OECD studies (OECD, 2022; OECD, 2024), is essential. Overall, the highest priority should be given to coordination and cooperation on space activities, establishing regulations, developing key technologies, and conducting measurement campaigns on launch emissions and reentry impacts (David et al., 2025).

### 4.4 A path to a truly sustainable governance of the global space commons

The accelerating growth of space activities leads to a nonlinear increase in risk and damage, as we have seen in previous sections. This makes it urgent to understand space as a *global commons*: a shared, finite domain whose benefits and vulnerabilities cross borders. No single nation or company can secure orbital safety on its own, and the consequences of irresponsible actions, such as space debris creation, atmospheric pollution, monopolization of orbital slots or retention of orbital data, affect all actors. Crucially, while risks and long-term costs are inherently collective, many of the most lucrative and environmentally impactful activities privatize gains. As space becomes more crowded and economically attractive, good governance must prevent overuse and ensure equitable access for present and future generations, based on outer space activities with the highest collective benefits.

The main space sustainability approaches, summarized in Table 4.4, differ in what they seek to protect, and in how they treat growth, substitution, responsibility, governance. They show the limitations of typical growth-compatible sustainability formulations, unless complemented by space governance as a shared commons respecting ecological limits and just wellbeing.

Weak sustainability and ESG/CSR are the framings most favored by current business practices, because they translate easily into risk management, disclosure, and "value creation". Strong sustainability, commons governance, and wellbeing-and-justice approaches impose hard limits, collective stewardship, and fair distribution of benefits and harms. If space is mainly treated as a source of profits, weak and managerial framings dominate; if it is treated as a global commons, business activity must be nested within collectively defined ecological, social, and intergenerational constraints.

Elinor Ostrom’s (1990) work on governing the commons shows the way different communities successfully managed such problems over long periods of time. Her research showed that commons can be managed sustainably when users collectively design and enforce rules that match the resource's ecological and social characteristics, even in highly complex settings. While her cases, from irrigation to fisheries and forests, differ from the outer space context, the underlying coordination problem is similar: collective benefits are maximized when rules are co-created, compliance is monitored, cheaters are punished, disputes are resolved transparently, and responsibilities are distributed across multiple levels of governance. Space governance is complicated by technological asymmetries, geopolitical competition, and the speed of commercial development. Nevertheless, Ostrom's principles offer a foundation for building a cooperative, polycentric system capable of adapting to uncertainty.

**Table 4.4:** Space governance models

| Space sustainability approach | Core idea / normative test | How growth is treated | Typical business / investor logic | Governance implications |
|---|---|---|---|---|
| **Strong sustainability** | Critical ecosystems and orbits must not be degraded beyond safe limits; multiple key natural functions are non-substitutable. | Growth is acceptable only within hard ecological and orbital boundaries. If growth undermines Earth systems, the orbital environment, or essential life-support functions, it is illegitimate. | Pushes firms away from "damage-offset-by-value" reasoning toward absolute limits, precaution, long time horizons, debris prevention, resource caps, and mission restraint. Requires business models compatible with sufficiency, durability, repair, reuse, and low externalization. | Requires binding international rules, enforceable limits, precautionary thresholds, monitoring, and liability. Supports treating key Earth-orbit-space relations as subject to planetary and intergenerational constraints, regardless of commercial implications. Absolute scale limitations. |
| **Weak sustainability** | Degradation of one form of capital may be acceptable if compensated by large gains in other forms of capital. | Growth and sustainability can be reconciled if overall value rises enough. Substitution is broadly allowed. | Fits mainstream economic and financial reasoning: environmental harm may be tolerated if outweighed by innovation, profits, connectivity, GDP, or strategic value. Encourages cost-benefit logic, offsets, risk pricing, and "net benefit" claims. No absolute protection. | Favors flexible regulation, disclosure, pricing of externalities, and managerial balancing. No hard caps. Governance mainly tries to correct market failures but does not limit expansion. |
| **Corporate sustainability / ESG / CSR** | Sustainability is integrated into firm strategy, reporting, legitimacy, and risk management. Emphasis is often on measurable environmental, social, and governance performance. | Usually assumes growth remains the primary objective, while harms are reduced, disclosed, mitigated, or reputationally managed. | Makes sustainability legible to firms and investors through compliance, reporting, stakeholder expectations, green finance, and risk reduction. Useful for incremental improvement. No absolute protection. Risk of using selective, sometimes misleading metrics. | Encourages standards, disclosure regimes, taxonomies, due diligence, and investor pressure. Helpful but insufficient where commons protection needs binding collective rules. |
| **Commons / Ostrom-inspired governance** | Space-related resources and environments are governed as shared commons through collectively defined rules, monitoring, graduated sanctions, conflict-resolution mechanisms, and participation by affected communities. | Growth is subordinate to maintaining the commons and ensuring fair, durable access. Expansion is legitimate only if it does not undermine collective use or long-term resource integrity. | Changes incentives from unilateral extraction and first-mover advantage toward stewardship, reciprocity, transparency, coordination, and restraint. Firms operate as users of a shared domain, subject to strict rules and monitoring. | Strongly supports polycentric governance, shared rule-making, monitoring, enforcement, liability, and institutional design for collective stewardship. This framing most directly supports the argument that space should be governed as a global commons. |
| **Wellbeing and justice** | The relevant test is whether space activity advances human and ecological wellbeing fairly: distributive justice, procedural justice, recognitional justice, as well as health, autonomy, and meaningful participation. | Growth is not an end in itself. Space activity is justified only if it contributes to equitable wellbeing without shifting harms onto vulnerable populations, future generations, or nonhuman systems. | Reorients business evaluation from profit to who benefits, who bears risks, who decides, and whether services genuinely support collective wellbeing. Challenges extractive or prestige-driven models and highlights inclusion, public value, and social legitimacy. | Requires participatory governance, fair benefit-sharing, representation of affected communities, attention to global inequality, and mechanisms to prevent appropriation of shared benefits by dominant firms or states. Similar to commons governance. |

Applying Ostrom's principles (numbered in square brackets) to space begins with **defining clear boundaries [1]** for resource use. Orbital space, launch rates, atmospheric pollution of launch and reentry, spectrum allocations, and access to limited resources like energy or metals, all require quantitative limits and transparent allocation rules. Without such boundaries, LEO congestion, debris cascades, atmospheric pollution and monopolization of critical orbits will critically increase. **Alignment with community needs [2]** requires that rules reflect the needs of diverse users, beyond dominant states or large companies. Essential services such as Navigation, Earth observation, and Scientific Exploration should be prioritized, while commercial deployments must not undermine equity, safety, or sustainability.

**Participatory decision-making [3]** is essential: all countries and major subnational communities, including those currently underrepresented, must have meaningful roles in shaping governing frameworks. This includes scientific communities, civil society, and Indigenous groups. **Monitoring and accountability [4]** must be independent and transparent, and must cover all main impacts: debris tracking, compliance with end-of-life requirements, atmospheric impacts (launches and re-entries), and orbital resources used by the large constellations.

To ensure compliance, **graduated sanctions [5]** are needed: minor violations addressed through warnings or increased reporting requirements, and serious infractions, such as negligent debris creation, unintended electromagnetic radiation (UEMR), or excessive spectrum use, penalized more strongly. **Conflict-resolution mechanisms [6]** should be fast, effective, and internationally legitimate, enabling cooperative responses to close approaches, contested resource claims, and signal interference. Finally, **nested governance networks [8]**, from national regulators to regional coalitions to global institutions, allow rules tailored to context while maintaining coherence across the entire system.[5]

The seven categories of space activity discussed in Section 3 help identify concrete governance priorities. For **scientific missions that expand understanding of the Universe**, impacts are relatively low and public benefits high. International research frameworks should ensure open data, equitable participation, and safeguards against future privatization of foundational findings. For activities contributing to **shared meaning and planetary identity**, such as symbolic missions and iconic imagery, governance should promote global representation and the cultivation of planetary citizenship and meaningful narratives.

---

[5] We did not include principle 7 (*commons need the right to organize*), as in the context of space sustainability, most mechanisms go beyond national law.

**Earth observation**, an enabling element of climate and biodiversity protection, requires fair access to data and collective decision-making on mission priorities. Establishing a global environmental data commons would help ensure that commercial EO growth does not erode the public's ability to respond to planetary crises. For **navigation and communication**, where resource use is rapidly escalating, governance must define orbital limits, prevent monopolization, ensure equitable access, and align constellation growth with both environmental and social needs. In **technology development**, transparency, open licensing, and priority for public-benefit innovation are essential to counter the tendency for private capture of gains and public absorption of environmental and social costs.

In the domain of **war, defense, and militarization**, cooperative governance is indispensable. Strengthening bans on space weapons and ASATs, establishing joint early-warning systems, and ensuring transparency can reduce the risk of escalation and accidental conflict. Finally, for **long-term futures**, such as planetary defense or deep-space habitation research, governance should ensure openness and avoid narratives that distract from urgent challenges on Earth. Narratives based on illusions of escape from Earth's problems are hugely damaging; research should contribute to global resilience.

Based on the activity-specific insights, we can define a broader pathway for achieving sustainable global space governance. We need three major changes. **First**, the international community must adopt a *strong sustainability* perspective: recognizing that the orbital environment has biophysical limits that cannot be substituted or offset. This implies setting thresholds for orbital use, emissions, and debris that align with long-term ecological stability. **Second**, governance must become *polycentric and inclusive* (as Ostrom emphasized) expanding beyond ITU, UNOOSA, and COPUOS to include bodies that can monitor and enforce rules while ensuring global meaningful participation, especially from the communities disproportionately affected by space activities. **Third**, the framing of outer space itself must shift. The perspective of space primarily as a frontier for endless economic expansion is highly problematic. It must be replaced by a vision of space as a shared life-support domain, a critical extension of Earth systems. Reframing space as a global commons enables governance arrangements that value long-term resilience.

While business perspectives often look for sustainability goals to be translated into incentives and market structures, our framework shows that such translation is itself a governance choice and must therefore be anchored in the treatment of space as a global commons. By developing these transformations based on Ostrom's principles and activity-specific recommendations, a truly sustainable governance becomes possible. It must protect the orbital and terrestrial environment, aligning space activities with the wellbeing of people and ecosystems.

## References

Abbas, S. (2025). *Challenges to space activities in the context of mega satellite constellations: A focus on environmental impacts.* Journal of Aerospace Studies and Science, 42(1), 1–18. https://www.janss.kr/archive/view_article?pid=jass-42-1-1

Adilov, N., Braun, V., Alexander, P., & Cunningham, B. (2023). An estimate of expected economic losses from satellite collisions with orbital debris. Journal of Space Safety Engineering, 10(1), 66–69.

Akhoondzadeh, M. (2024). Earthquake prediction using satellite data: Advances and ahead challenges. Advances in Space Research, 74(8), 3539–3555. DOI: 10.1016/j.asr.2024.06.054

Alexandrow, C. (2008). The Story of GPS. In DARPA: 50 Years of Bridging the Gap (pp. 54-55). Defense Advanced Research Projects Agency.

Al-Hraishawi, H., Chougrani, H., Kisseleff, S., Lagunas, E., & Chatzinotas, S. (2023). A survey on nongeostationary satellite systems: The communication perspective. IEEE Communications Surveys & Tutorials, 25(1), 101–132. DOI: 10.1109/COMST.2022.3197695

Anderson, S. J., Kubiszewski, I., & Sutton, P. C. (2024). *The ecological economics of light pollution: Impacts on ecosystem service value*. Remote Sensing, 16(14), 2591. DOI: 10.3390/rs16142591

Barentine, J.C. et al., 2023, "Aggregate Effects of Proliferating LEO Objects and Implications for Astronomical Data Lost in the Noise" https://arxiv.org/pdf/2302.00769

Bassa, C. G., Aartsen, M. G., Alexov, A., Anderson, J., Anderson, J. M., et al. (2024). *Bright unintended electromagnetic radiation from second-generation Starlink satellites*. Astronomy & Astrophysics, 687, A56. DOI: 10.1051/0004-6361/202451856

Bassa, C. G., Hainaut, O. R., & Galadí-Enríquez, D. (2022). Analytical simulations of the effect of satellite constellations on optical and near-infrared observations. *Astronomy & Astrophysics, 657*, A75. DOI: 10.1051/0004-6361/202142101

Bauer, M. E., Cipra, J. E., Anuta, P. E., & Etheridge, J. B. (1979). Identification and area estimation of agricultural crops by computer classification of Landsat MSS data. Remote Sensing of Environment, 8(1), 77-92.

Butash, T., Garland, P., & Evans, B. G. (2021). Non-geostationary satellite orbit communications satellite constellations history. International Journal of Satellite Communications and Networking, 39(1), 1–5. https://doi.org/10.1002/sat.1375

Bongers, A., & Torres, J. L. (2023). Orbital debris and the market for satellites. Ecological Economics, 209, 107831. DOI: 10.1016/j.ecolecon.2023.107831

Bonnal, C., McKnight, D., Jorgenson, C., & Masson-Zwaan, T. (2025). IAF–IISL–IAA initiative on space traffic management: Synthesis report on IAF technical committee TC 26 on space traffic management. Acta Astronautica, 232, 706–720. DOI: 10.1016/j.actaastro.2025.03.024

Bridging the digital divide: A data-driven approach to transformational connectivity. (2025). International Telecommunication Union. https://www.itu.int/hub/publication/d-inno-whitepaper-2025/

ClearSpace-1. (n.d.). Retrieved from https://clearspace.today/missions/clearspace-1

Colombo, C., Martinez, P., Letizia, F., Cattani, B., David, E., Escobar, D., Kawamoto, S., Rovetto, R., & Sumelzo, I. (2025). Space capacity management and its interaction with space traffic management. Acta Astronautica. DOI: 10.1016/j.actaastro.2025.01.069

Committee on Radio Astronomy Frequencies. (2025). *About CRAF.* European Science Foundation. https://www.craf.eu/about-craf/

COPUOS. (2025). *Report of the Committee on the Peaceful Uses of Outer Space: 68th session, 25 June–2 July 2025* (A/80/20). United Nations Office for Outer Space Affairs.

Committee on Space Research. (2026, June 30). COSPAR policy on planetary protection. https://cosparhq.cnes.fr/cospar-policy-on-planetary-protection/

David, E., Burger, E., Finnlay, C., Guyot, B., Hellmich, S., Kleint, L., Losch, A., Mueller, D., Poirier, C., Treyer, K., Udriot, M., & Verkammen, M. (2025). Swiss Space Sustainability Research Days 2025: Outcomes and strategic roadmap.

David, E., Hellmich, S., & Crichton, C. (2025). *The importance of dark and quiet skies in Swiss astronomy and the dialog with the space industry*

Di Vruno, F., Winkel, B., Bassa, C. G., Józsa, G. I. G., Brentjens, M. A., Jessner, A., & Garrington, S. (2023). *Unintended electromagnetic radiation from Starlink satellites detected with LOFAR between 110 and 188 MHz.* arXiv.

Dominguez Calabuig, G. J., Miraux, L., Wilson, A. R., Sarritzu, A., & Pasini, A. (2022). Eco-design of future reusable launchers: Insight into their life cycle and atmospheric impact. In Proceedings of the 9th European Conference for Aeronautics and Space Sciences. https://doi.org/10.13009/EUCASS2022-7353

Edgington, S. G., & Spilker, L. J. (2016). Cassini's grand finale. Nature Geoscience, 9, 472–473. https://doi.org/10.1038/ngeo2753

Engelbrecht, B. N., et al. (2024). *Radio frequency interference from radio navigation satellite systems: Simulations and comparison to MeerKAT single-dish data.* Monthly Notices of the Royal Astronomical Society. DOI: 10.1093/mnras/stae2649

ESA LCA Working Group. (2016). Space system Life Cycle Assessment (LCA) guidelines. European Space Agency.

Estes, J. E., Jensen, J. R., & Tinney, L. R. (1978). Remote sensing of agricultural water demand information: A California study. Water Resources Research, 14(2), 170-176.

European Cooperation for Space Standardization. (2024). ECSS-U-AS-10C Rev. 2 – Space debris mitigation. ECSS Secretariat. https://ecss.nl/standards/ecss-u-as-10c-rev-2-space-debris-mitigation/

European Space Agency, Space Debris Office. (2025). ESA Space Environment Report 2025.https://www.sdo.esoc.esa.int/publications/Space_Environment_Report_I9R1_20251021.pdf

European Space Agency. (2023a). ESA Space Debris Mitigation Requirements (ESSB-ST-U-007, Issue 1, 30 Oct 2023). ESA. https://technology.esa.int/upload/media/ESA-Space-Debris-Mitigation-Requirements-ESSB-ST-U-007-Issue1.pdf

European Space Agency. (2023b). Zero Debris Charter: Towards a safe and sustainable space environment. https://esoc.esa.int/sites/default/files/Zero_Debris_Charter_EN.pdf

Fabio Falchi et al. ,The new world atlas of artificial night sky brightness. Sci. Adv.2,e1600377(2016).DOI:10.1126/sciadv.1600377

Falchi, F., Cinzano, P., Duriscoe, D., Kyba, C. C. M., Elvidge, C. D., Baugh, K., Portnov, B. A., Rybnikova, N. A., & Furgoni, R. (2016). *The new world atlas of artificial night sky brightness. Science Advances, 2*(6), e1600377.

Ferreira, J. P., Huang, Z., Nomura, K. I., & Wang, J. (2024). Potential ozone depletion from satellite demise during atmospheric reentry in the era of mega-constellations. Geophysical Research Letters, 51(11), e2024GL109280.

Ferreira, J. P., Nomura, K., & Wang, J. (2023, September). Preliminary assessment of the environmental impact of space debris demise during atmospheric reentry. In Advanced Maui Optical and Space Surveillance Technologies (AMOS) Conference.

Finlay, C., Bassett, B. A., Kunz, M., & Oozeer, N. (2023). *Trajectory-based RFI subtraction and calibration for radio interferometry.* Monthly Notices of the Royal Astronomical Society. DOI: 10.1093/mnras/stad1979

Fischer, J. S., Fasoulas, S., Bergmann, N., Ott, A., Pinto, V., Wilken, J., ... & Maddock, C. (2024, November). A pathway for closing the knowledge gaps for a comprehensive life cycle assessment and ecodesign of space transportation systems: results of the 3rd Workshop on Life Cycle Assessment of Space Transportation Systems. In *Workshop on Life Cycle Assessment of Space Transportation Systems*. DOI: 10.5281/zenodo.14222106

Fischer, J.-S., & Fasoulas, S. (2024). *Assessment of launch and re-entry emissions of space transportation systems and their environmental impact*. DOI: 10.52202/078373-0079

Guardabasso, P., Lizy-Destrez, S., & Ansart, M. (2021). Lunar orbital debris mitigation: Characterisation of the environment and identification of disposal strategies. In T. Flohrer, S. Lemmens, & F. Schmitz (Eds.), Proceedings of the 8th European Conference on Space Debris. ESA Space Debris Office. https://conference.sdo.esoc.esa.int/proceedings/sdc8/paper/42/SDC8-paper42.pdf

Hainaut, O. R., & Williams, A. P. (2020). Impact of satellite constellations on astronomical observations with ESO telescopes in the visible and infrared domains. Astronomy & Astrophysics, 636, A121. https://doi.org/10.1051/0004-6361/202037501

Hamacher, D. W., De Napoli, K., & Mott, B. (2020). *Whitening the sky: Light pollution as a form of cultural genocide. Journal of Dark Sky Studies*, 1, 1–15. https://arxiv.org/abs/2001.11527

Houborg, R., & McCabe, M. F. (2016). High-Resolution NDVI from planet’s constellation of earth observing nano-satellites: A new data source for precision agriculture. Remote Sensing, 8(9), 768.

Hull, S. M. (2013). End of Mission considerations: Chapter 30. Disposal guidelines and requirements (NASA/TP-2013-4016, pp. 30.1-30.6). NASA Goddard Space Flight Center. https://ntrs.nasa.gov/api/citations/20130000278/downloads/20130000278.pdf?

IAU Centre for the Protection of the Dark and Quiet Sky from Satellite Constellation Interference (IAU CPS). (2023, January 10). NSF and SpaceX sign agreement to mitigate impact of Starlink satellites on ground-based astronomy. IAU CPS.

International Astronomical Union Centre for the Protection of the Dark and Quiet Sky from Satellite Constellation Interference. (2024). *Call to protect the dark and quiet sky from harmful interference by satellite constellations.* https://cps.iau.org/documents/49/techdoc102.pdf

International Astronomical Union, IAU Centre for the Protection of the Dark and Quiet Sky. (2020). *Dark and Quiet Skies for Science and Society: Report and recommendations.*

International Organization for Standardization. (2006a). Environmental management — Life cycle assessment — Principles and framework (ISO 14040:2006). ISO.

International Organization for Standardization. (2006b). Environmental management — Life cycle assessment — Requirements and guidelines (ISO 14044:2006). ISO.

International Organization for Standardization. (2023). ISO 24113:2023 – Space systems: Space debris mitigation requirements. ISO. https://www.iso.org/standard/81287.html

International Telecommunication Union. (2022). ITU Facts and Figures 2022: The Gender Digital Divide. International Telecommunication Union. https://www.itu.int/hub/publication/d-ind-ff-2022-gender/

International Telecommunication Union. (n.d.). About ITU. International Telecommunication Union. https://www.itu.int/en/about/

International Telecommunication Union Radiocommunication Sector. (2025). *Managing the radio frequency spectrum.*

International Telecommunication Union. (2010). Environmental protection of the geostationary-satellite orbit (Recommendation ITU-R S.1003-2). https://www.itu.int/dms_pubrec/itu-r/rec/s/R-REC-S.1003-2-201012-I!!PDF-E.pdf

IPCC, Lee, H., Calvin, K., Dasgupta, D., Krinner, G., Mukherji, A., Thorne, P., ... & Zommers, Z. (2023). Climate change 2023: synthesis report. Contribution of working groups I, II and III to the sixth assessment report of the intergovernmental panel on climate change.

IPCC, 2022: Summary for Policymakers [H.-O. Pörtner, D.C. Roberts, E.S. Poloczanska, K. Mintenbeck, M. Tignor, A. Alegría, M. Craig, S. Langsdorf, S. Löschke, V. Möller, A. Okem (eds.)]. In: Climate Change 2022: Impacts, Adaptation, and Vulnerability. Contribution of Working Group II to the Sixth Assessment Report of the Intergovernmental Panel on Climate Change [H.-O. Pörtner, D.C. Roberts, M. Tignor, E.S. Poloczanska, K. Mintenbeck, A. Alegría, M. Craig, S. Langsdorf, S. Löschke, V. Möller, A. Okem, B. Rama (eds.)]. Cambridge University Press, Cambridge, UK and New York, NY, USA, pp. 3-33, doi:10.1017/9781009325844.001

Jakhu, R. S. (2010). Iridium-Cosmos collision and its implications for space operations. In Yearbook on Space Policy 2008/2009: Setting New Trends (pp. 254-275). Springer Vienna.

Kapoor, T. S., Upadhyay, P., Huang, J., Ren, G., Cavin, J., Mitroo, D., ... & Chakrabarty, R. K. (2025). Shortwave Absorption by Alumina Aerosol Amplifies Stratospheric Warming. arXiv preprint arXiv:2509.24246.

Kazansky, Y., Wood, D., & Sutherlun, J. (2016). The current and potential role of satellite remote sensing in the campaign against malaria. Acta Astronautica, 121, 292-305.

Kessler, D. J., & Cour-Palais, B. G. (1978). Collision frequency of artificial satellites: The creation of a debris belt. Journal of Geophysical Research: Space Physics, 83(A6), 2637–2646. DOI: 10.1029/JA083iA06p02637

Kirkpatrick, C. R. (2024). A Technical and Policy Needs Analysis for Space Traffic Management of Low Lunar Orbit (Doctoral dissertation, Massachusetts Institute of Technology).

Koller, J. S., & Thompson, A. (2020). *Light pollution from satellites.* Centre for Space Policy and Strategy.

Koller-Thompson, A. (2020). Light pollution in space. Center for Space Policy and Strategy.https://csps.aerospace.org/sites/default/files/2021-08/Koller-Thompson_LightPollution_20201026.pdf

Kyek, S. J. (2023). *Light pollution as an environmental issue: Governance and regulatory approaches in Switzerland* [Master's thesis, ETH Zurich]. Swiss Federal Institute for Forest Snow and Landscape Research.

Liou, J.-C., & Shoots, D. (2009, July). United Nations' COPUOS receives update on Iridium-Cosmos collision. Orbital Debris Quarterly News, 13(3), 2. National Aeronautics and Space Administration. https://ntrs.nasa.gov/api/citations/20090027660/downloads/20090027660.pdf

Lloyd, W. F. (1833). Two lectures on the checks to population: Delivered before the University of Oxford, in Michaelmas Term 1832. S. Collingwood.

Locke, J., Colvin, T. J., Ratliff, L., Abdul-Hamid, A., & Samples, C. (2024). Cost and benefit analysis of mitigating, tracking, and remediating orbital debris. Cost and Benefit Analysis of Mitigating, Tracking, and Remediating Orbital Debris.

Lovell, A. C. B., Blackwell, M. R., & Wilson, R. (1962). West Ford Project, Interference to Astronomy from Belts of Orbiting Dipoles (Needles). Quarterly Journal of the Royal Astronomical Society, 3, 100-108.

Maloney, C. M., Portmann, R. W., Ross, M. N., & Rosenlof, K. H. (2025). Investigating the potential atmospheric accumulation and radiative impact of the coming increase in satellite reentry frequency. Journal of Geophysical Research: Atmospheres, 130(6), e2024JD042442.

Marín, C. (2011). Starlight: A common heritage. IAU Symposium 260, 449 to 456. https://adsabs.harvard.edu/pdf/2011IAUS..260..449M

McDonald, D. (2025). Preserving cosmic heritage: The cultural significance of astronomy and the imperative for dark and quiet skies protection in outer space.

Miraux, L. (2022). Environmental limits to the space sector's growth. Science of the Total Environment, 806, 150862. https://doi.org/10.1016/j.scitotenv.2021.150862

Mohanty, U. C., Osuri, K. K., Tallapragada, V., Marks, F. D., Pattanayak, S., Mohapatra, M., ... & Niyogi, D. (2015). A great escape from the Bay of Bengal "super sapphire–Phailin" tropical cyclone: a case of improved weather forecast and societal response for disaster mitigation. Earth Interactions, 19(17), 1-11.

Murphy, D. M., Abou-Ghanem, M., Cziczo, D. J., Froyd, K. D., Jacquot, J., Lawler, M. J., ... & Shen, X. (2023). Metals from spacecraft reentry in stratospheric aerosol particles. Proceedings of the National Academy of Sciences, 120(43), e2313374120.

NASA. (2012). NASA Technical Standard: Process for Limiting Orbital Debris (NASA-STD-8719.14A); Requirement 56626, p.44. https://soma.larc.nasa.gov/SIMPLEx/pdf_files/871914.pdf

NASA. (2014). *NASA Radio Frequency Spectrum Management Manual (NPR 2570.1C).* https://nodis3.gsfc.nasa.gov/displayDir.cfm?c=2570&s=1C

NASA. (2017). Micrometeoroid and orbital debris (MMOD) handbook (NASA/SP-2017-614). National Aeronautics and Space Administration.

NASA. (2019). NASA-STD-8719.14C: Process for limiting orbital debris. NASA Technical Standards. https://standards.nasa.gov/standard/nasa/nasa-std-871914c

NASA. (2024). Where are Voyager 1 and Voyager 2 now? https://science.nasa.gov/mission/voyager/where-are-voyager-1-and-voyager-2-now/

National Academies of Sciences, Engineering, and Medicine. (2017). Introduction (In Powering Science: NASA's large strategic science missions). The National Academies Press. DOI: 10.17226/24857

NEC XON and Skywire deploy OneWeb satellite solution in South Africa. (2025, January 27). Satellite Today. https://www.satellitetoday.com/connectivity/2025/01/27/nec-xon-and-skywire-deploy-oneweb-satellite-solution-in-south-africa/

Nozawa, W., Kurita, K., Tamaki, T., & Managi, S. (2023). To what extent will space debris impact the economy? Space Policy, 66, 101580. DOI: 10.1016/j.spacepol.2023.101580

NSR. (2022, April 20). NSR: Developing Moon market propelled by 250+ missions and $105 billion in revenue through decade. GlobeNewswire. https://www.globenewswire.com/en/news-release/2022/04/20/2425653/0/en/NSR-Developing-Moon-Market-Propelled-by-250-Missions-and-105-Billion-in-Revenue-through-Decade.html

OECD. (2022). Earth's Orbits at Risk: The Economics of Space Sustainability. OECD Publishing. DOI: 10.1787/16543990-en

OECD. (2024). The Economics of Space Sustainability. OECD Publishing. DOI: 10.1787/b2257346-en

Osborn, J., Otarola, A., Blake, C., Butterley, T., Saunter, C., Wilson, R. W., & Massey, R. (2022). Astrosat: Forecasting satellite transits for optical astronomical observatories. *Monthly Notices of the Royal Astronomical Society, 509*(2), 1848–1863. https://academic.oup.com/mnras/article/509/2/1848/6406501

Ostrom, E. (1990). Governing the commons: The evolution of institutions for collective action (Vol. 10). Cambridge: Cambridge university press.

Ott, A., & Bonnal, C. (2025). Atmospheric re-entry of orbital objects–can "Design For Non-Demise" be the optimal solution?. Acta Astronautica.

Poole, R. (2008). Earthrise: How man first saw the Earth. Yale University Press. (ISBN 0300137664)

Pardini, C., & Anselmo, L. (2017). Revisiting the collision risk with cataloged objects for the Iridium and COSMO-SkyMed satellite constellations. Acta Astronautica, 134, 23-32.

Parmar, A., Bonnet, R. M., De Marchi, G., García-Lario, P., Kuulkers, E., Pilbratt, G., ... & Zarnecki, J. (2025). ESA Science Programme Missions: Contributions and Exploitation (p. 192). Springer Nature.

Pleney, L. (2025, January 20). Highlights of the 2024 space economy. Novaspace. https://nova.space/in-the-loop/highlights-of-the-2024-space-economy/

Rani, S. I., Prasad, V. S., Rajagopal, E. N., & Basu, S. (2016, May). Height of warm core in very severe cyclonic storms Phailin: INSAT-3D perspective. In Remote Sensing and Modeling of the Atmosphere, Oceans, and Interactions VI (Vol. 9882, pp. 224-229). SPIE.

Rao, A., & Rondina, G. (2025). The economics of orbit use: Open access, external costs, and runaway debris growth. Journal of the Association of Environmental and Resource Economists, 12(2), 353–388. DOI: 10.1086/730356

Ross, M. N., & Sheaffer, P. M. (2014). Radiative forcing caused by rocket engine emissions. Earth's Future, 2(4), 177–196. https://doi.org/10.1002/2013EF000160

Ross, M. N., & Jones, K. L. (2022). Implications of a growing spaceflight industry: Climate change. Journal of Space Safety Engineering, 9(3), 469-477.

Ross, M., & Vedda, J. A. (2018). The policy and science of rocket emissions. Center for Space Policy and Strategy, The Aerospace Corporation.

Rossi, A., Sánchez-Ortiz, N., David, E., Opromolla, R., & Grishko, D. (2024). Future activities in the near-earth space in the face of ever-increasing space traffic. Acta Astronautica, 225, 891–897.DOI: 10.1016/j.actaastro.2024.09.063

Ryan, R. G., Marais, E. A., Balhatchet, C. J., & Eastham, S. D. (2022). Impact of rocket launch and space debris air pollutant emissions on stratospheric ozone and global climate. Earth's Future, 10(6), e2021EF002612.

Sagan, C., Druyan, A., & Soter, S. (Writers), & Kennard, D., Malone, A., & McCain, R. (Directors). (1980, November 23). The lives of the stars (Season 1, Episode 9) [TV series episode]. In A. Malone (Executive Producer), Cosmos: A personal voyage. Carl Sagan Productions; KCET.

Schulz, L., Glassmeier, K. H., Herberhold, M., Mitchell, A., Murphy, D. M., Plane, J., & Plaschke, F. (2025). Space waste: An update of the anthropogenic matter injection into Earth atmosphere. arXiv preprint arXiv:2510.21328.

Sector, D. (2025). Measuring digital development. https://www.itu.int/hub/publication/D-IND-ICT_MDD-2025-3/

Shahar, K., & Greenbaum, D. (2020). Lessons in space regulations from the lunar tardigrades of the Beresheet hard landing. Nature Astronomy, 4(3), 208–209. https://doi.org/10.1038/s41550-020-1034-2

SKA Observatory. (2025). *Frequently asked questions: How much will the SKA telescopes cost?* SKA Observatory. https://www.skao.int/en/resources/291/frequently-asked-questions

Stoppa, F., Bonnarel, F., Louys, M., Mantelet, G., & Molinaro, M. (2024). Automated detection of satellite trails in ground-based optical observations using deep learning. *Astronomy & Astrophysics, 692*, A207.

Stoppa, F., Groot, P. J., Stuik, R., Vreeswijk, P., Bloemen, S., Pieterse, D. L. A., & Woudt, P. A. (2024). Automated detection of satellite trails in ground-based observations using U-Net and Hough transform. Astronomy & Astrophysics, 692, A199.

Sun, Q., & Liu, H. (2025). Space safety in the age of LEO constellations: The role of spectrum management. Journal of Space Safety Engineering. 10.1016/j.jsse.2025.07.001

Summit of the Future Outcome Documents: Pact for the Future, Global Digital Compact and Declaration on Future Generations. (2024). United Nations. https:www.un.org/sites/un2.un.org/files/sotf-pact_for_the_future_adopted.pdf

The Harvard Crimson. (1963, May 24). Project West Ford — "Four hundred million copper wires are now circling the globe…" The Crimson.

Tormena, E., Soares, T., Serrano, S., Affentranger, L., Padilla, E., & Turchetto, T. (2024). A simplified approach to LCA for space systems. DOI: 10.52202/078372-0056

Tucker, C. J., Vanpraet, C., Boerwinkel, E., & Gaston, E. A. (1983). Satellite remote sensing of total dry matter production in the Senegalese Sahel. Remote Sensing of Environment, 13(6), 461-474.

Turchetto, T., & Guardabasso, P. (2024). Space debris mitigation guidelines for lunar orbits. In Space debris mitigation guidelines for lunar orbits. https://doi.org/10.5075/epfl.20.500.14299/240544

Udriot, M., Treyer, K., Verkammen, M., & Fischer, J.-S. (2025). *REACT announcement of opportunity* (Version v1) [White paper]. Zenodo. https://doi.org/10.5281/zenodo.14802424

United Nations Office for Outer Space Affairs. (2010). Space debris mitigation guidelines of the Committee on the Peaceful Uses of Outer Space (A/AC.105/890). United Nations.

United Nations Office for Outer Space Affairs. (2021). Guidelines for the Long-term Sustainability of Outer Space Activities of the Committee on the Peaceful Uses of Outer Space. DOI: 10.18356/9789210021852

United Nations Office for Outer Space Affairs. (2025a). Benefits of space: Disasters. https://www.unoosa.org/oosa/de/benefits-of-space/disasters.html

United Nations Office for Outer Space Affairs (UNOOSA). (2025b, February 4). Conference Room Paper on the Protection of Dark and Quiet Skies for Science and Society (A/AC.105/C.1/2025/CRP.22/Rev.2). United Nations Committee on the Peaceful Uses of Outer Space.

Utah Community Development Office. (2020). *Guidance & best practices: Dark sky planning. An introduction for local leaders.* State of Utah Department of Workforce Services, Housing & Community Development Office.

Voigt, C., Schumann, U., Graf, K., & Gottschaldt, K.-D. (2013). Impact of rocket exhaust plumes on atmospheric composition and climate—An overview. Progress in Propulsion Physics, 4, 657–670. https://doi.org/10.1051/eucass/201304657

Wilson, A. R., & Vasile, M. (2023). The space sustainability paradox. Journal of Cleaner Production, 423, 138869.

Wood, D., Rathnasabapathy, M., Stober, K. J., & Menon, P. (2024). Challenges and progress in applying space technology in support of the sustainable development goals. Acta Astronautica, 219, 678–692. https://doi.org/10.1016/j.actaastro.2024.03.064

World Economic Forum. (2023). WEF State of the Connected World 2023 Edition. https://www3.weforum.org/docs/WEF_State_of_the_Connected_World_2023_Edition.pdf

World Economic Forum. (2025). WEF Global Risks Report 2025. https://www.weforum.org/publications/global-risks-report-2025/

World Meteorological Organization & United Nations Environment Programme. (2022). Scientific Assessment of Ozone Depletion: 2022 (GAW Report No. 278).

Yavuz, C., & Bektas, S. (2024). *The negative impacts of light pollution caused by satellite mega-constellations on the environment and the approach of international space law to the issue.* International Journal for Research in Applied Science & Engineering Technology, 12(12). DOI: 10.22214/ijraset.2024.65957